\documentclass[pra,10pt,twocolumn,floatfix]{revtex4-1} %%

\usepackage {graphicx} %%
\usepackage {amssymb} %%
\usepackage {amsmath}

\newcommand{\reffig}[1]{Fig.~\ref{#1}}
\newcommand{\refeq}[1]{Eq.~(\ref{#1})}

\begin{document}

%\begin{frontmatter}

\title{Self-starting stable coherent mode-locking in a two-section laser}

%% Group authors per affiliation:
\author{R. M. Arkhipov$^{a}$, M. V. Arkhipov$^b$, I. Babushkin$^{c,d}$}
\address{$^a$ ITMO University,  Kronverkskiy prospekt, 49, 197101 St. Petersburg, Russia,   \\
  $^b$ Faculty of Physics, St. Petersburg State University, Ulyanovskaya 1, Petrodvoretz, St. Petersburg 198504, Russia \\
  $^c$ Institute of Quantum Optics, Leibniz University Hannover,
  Welfengarten 1 30167, Hannover, Germany \\
  $^d$ Max Born Institute, Max Born Str. 2a, 12489 Berlin, Germany}

\begin{abstract}
  Coherent mode-locking (CML) uses self-induced transparency (SIT)
  soliton formation to achieve, in contrast to conventional schemes
  based on absorption saturation, the pulse durations below the limit
  allowed by the gain line width. Despite of the great promise it is
  difficult to realize it experimentally because a complicated setup
  is required. In all previous theoretical considerations CML is
  believed to be non-self-starting.  In this article we show that if
  the cavity length is selected properly, a very stable (CML) regime
  can be realized in an elementary two-section ring-cavity geometry,
  and this regime is self-developing from the non-lasing state. The
  stability of the pulsed regime is the result of a dynamical
  stabilization mechanism arising due to finite-cavity-size effects.
\end{abstract} 

%\begin{keyword}
%\texttt{elsarticle.cls}\sep \LaTeX\sep Elsevier \sep template
%\MSC[2010] 00-01\sep  99-00
%\end{keyword}

%\end{frontmatter}

%\linenumbers


\maketitle

\section{Introduction}

Development of ultrashort laser pulse sources with high repetition
rates and peak power is an area of principal interest in optics.  Such
lasers have applications in a high-bit-rate optical communications,
real time-monitoring of ultrafast processes in matter etc.  A
well-known method for generating high power ultrashort optical pulses
is a passive mode-locking (PML)
\cite{haus2000mode,Khanin,RafailovNat,Rafailov,Kryukov,arsenijevicpassive}.
In order to achieve PML, a nonlinear saturable absorbing medium is
placed into the laser cavity. In most of existing passively
mode-locked lasers generation of ultrashort pulses arises due to the
absorption/gain saturation in the gain and absorber sections
\cite{haus1975theory,kartner1996soliton,
  kurtner1998mode,haus2000mode}, with only few exceptions, such as in
the case Kerr-lens mode-locking. Thus, in the most schemes the
ultimate limit on the pulse duration $\tau_p$ is set by the medium
polarization relaxation time $T_2$, that is, $\tau_p\gtrsim T_2$. The
interaction of the pulse with resonant gain and absorber media is not
coherent in the sense that the medium polarization just follows the
field and thus can be adiabatically eliminated\cite{haus1975theory,
  haus1975, new1974pulse,kartner1996soliton,
  kurtner1998mode,haus2000mode,vladimirov2005,arkhipov2013hybrid,arkhipov2015pulse}. This
is valid also in the case when the absorber enters the coherent regime
(see below) whereas the gain medium is still in the usual regime of
the gain saturation
\cite{vasilev86,kalosha99a,baer12,harvey94,talukder14}.

\begin{figure}[htbp!]
\begin{centering}
\includegraphics[width=0.5\textwidth]{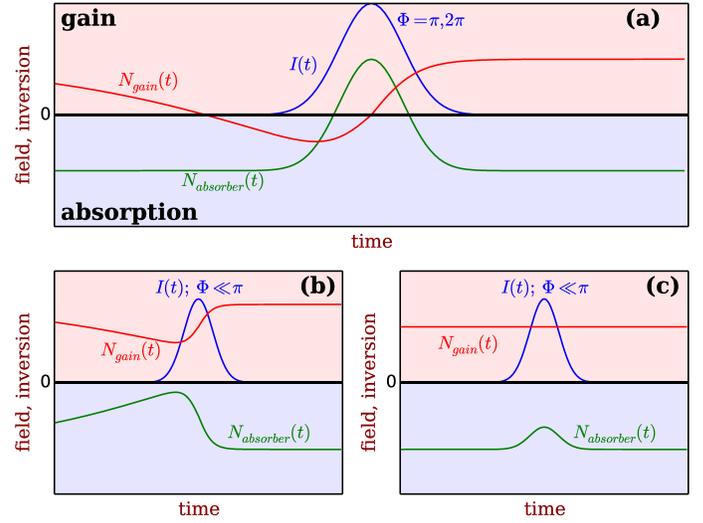}%%
\caption{ \label{fig:0} The essential details of coherent mode-locking
  (CML) (sometimes refereed to as ``SIT mode-locking''). (a) The
  dependence of the field intensity $I(t)$ and population inversion
  $N_{gain}$, $N_{absorber}$ in the gain and absorber sections in the
  CML regime. The pulse area $\Phi$ [see \refeq{eq:phi}] is $\pi$ in
  the gain section (a half of Rabi oscillation) and $2\pi$ in the
  absorber (a whole Rabi oscillation); in both sections the population
  inversion changes its sign. In the absorber, the population returns
  back on the pulse duration despite of much slower relaxation time
  $T_{1a}$ of the absorber medium. The gain section is also switched
  from ``amplifying'' to ``absorbing'' state and then slowly, on the
  time scale of the population relaxation time $T_{1g}$, returns
  back. For comparison, in (b) passive mode-locking with a slow
  saturable absorber and gain saturation is demonstrated. In this
  case, both gain and absorber sections are only saturated, without
  onset of Rabi oscillations ($\Phi\ll\pi$). In this case, long tails
  of population inversion relaxing after the pulse passage appear. In
  (c), mode-locking with a fast absorber (such as Kerr lens) and no
  gain saturation is presented. Note some similarity of the absorber
  behavior in (a) and (c) despite of fundamentally different
  relaxation rates of the absorber [$T_{1a}$ in (a) and instantaneous
  in (c)], which is the consequence of the coherent regime.  }
\end{centering} 
\end{figure}

Another new way to achieve ultrashort pulse generation was proposed
theoretically in \cite{kozlov1997self,kozlov1999self} [see
\reffig{fig:0}(a)] where it was named ``coherent mode-locking'' (CML),
or self-induced-transparency (SIT) mode-locking, as it is called
sometimes \cite{menyuk2009self,talukder2009analytical}. In this
approach interaction of light with matter is so strong, that the
medium polarization and inversion change significantly on the time
scale of the pulse duration, and Rabi oscillations arise
\cite{mccall1969self,kryukov70,Allen:book} (coherent regime).  In this
case, because the period of Rabi oscillations is not limited from
below, the pulse duration can be significantly smaller then the medium
coherence time, $\tau_p\ll T_2$, and the presence of the phase memory
on the scale of $T_2$ changes the evolution of the pulse dramatically.
Unlike the common passive mode-locking with a slow saturable absorber
\cite{haus1975theory,haus2000mode,dudling:book,keller10}, where the
absorption is just saturated near the pulse center [see
\reffig{fig:0}(b)], in the case of coherent interaction it is
completely inverted, so that the population inversion crosses zero and
changes its sign. As a result, such pulse propagates without losses in
the absorber in the regime of self-induced-transparency (SIT) ($2\pi$
pulse), thus forming a soliton. In the gain section the pulse takes
all the energy from the medium ($\pi$ pulse), making it highly
absorbing (that is, again, the population inversion changes it sign),
in contrast to common lasers schemes, were the population inversion in
the gain section either does not change significantly [as in
\reffig{fig:0}(c)], or changes relatively slightly, without crossing
zero [as in \reffig{fig:0}(b)].  Because of this, such SIT-induced
solitons are fundamentally different in their dynamics from the pulses
appearing in the saturable schemes
%, even if the later also can be
%considered as localized structures
\cite{mollenauer84,dudling:book,keller10}.
 
Contrary to the conventional passively mode-locked lasers with a
saturable absorber, CML lasers can generate optical pulses with a
duration much shorter than $T_2$, i.e. with the spectrum exceeding the
bandwidth of the gain medium. Moreover, it was predicted
\cite{kozlov1997self,kozlov1999self,
  kozlov2011obtaining,kozlov2013single} that pulse duration from CML
lasers can approach the single optical cycle limit despite of
narrow-band gain and absorber, even in the presence of inhomogeneous
line broadening, dispersion, and the complex level structure
\cite{vysotina09}.

Unfortunately, despite of the great promise, there was no experimental
demonstration of CML in configuration proposed in
\cite{kozlov1997self,kozlov1999self,
  kozlov2011obtaining,kozlov2013single} up to now.  However,
generating of short pulses shorter than $T_2$ in mode-locked argon-ion
laser with active mode-locker and in self-locked He-Ne laser was
demonstrated experimentally in Refs. \cite{harvey1989superfluorescent,
  dudley1993coherent} and \cite{lis1994self}, respectively.  In our
recent work \cite{arkhipov2015mode} we have also shown experimentally
a mode-locking regime with a pulse duration less than $T_2$ in the
absorber.
Also, in \cite{menyuk2009self,talukder2009analytical} quantum cascade
laser structures were proposed as candidates for experimental
realization of CML regime.  Theoretical study of CML performed in
\cite{kozlov1997self,kozlov1999self,
  kozlov2011obtaining,kozlov2013single,menyuk2009self,talukder2009analytical}
was carried out for a laser with the gain and absorber implemented
within the same sample, that is, as a ``homogeneous mixture'' of the
amplifying and absorbing media. %Such ``mixed''
  Such proposals have, however, some important disadvantages.  First,
  the pulsed regime can not develop spontaneously from a non-lasing
  state. Namely, to ensure the stability of the pulse one have to
  suppress the background fluctuations far away from it, thus
  automatically making the non-lasing steady state stable and the
  whole laser non-self-starting. That is, to initiate a soliton, one
  needs a seed pump pulse injected to the laser. The necessity to make
  CML lasers non-self-starting follows thus from a solitonic character
  of the coherent mode-locking. In contrast, many non-CML mode-locking
  lasers can indeed stably self-start even if the absorber (but not
  the amplifier) works in the coherent regime
  \cite{kalosha99a,talukder14}.
  Nevertheless the problem of
    self-starting is also actual for other existing fast
      modelocking schemes such as Kerr-lens
    mode-locking \cite{Kryukov}. The second
  important drawback is the necessity of a ``homogeneous mixture'' of
  the amplifier and absorber assumed in the works on CML up to now,
  which, although ensures a solitonic character of the mode-locking,
  makes its practical implementation rather difficult.
  
  In this article, we consider a simple scheme of CML-based
  modelocking with the amplifying and absorbing media being well
  separated spatially, forming rather usual two-section geometry. A
  possibility of the CML is this case was shortly reported by us in
  \cite{arkhipov2015coherent}. In the present article we focus on the
  problem of self-starting of CML regime in such geometry and analyze
  in details characteristic behaviour of the system.  We
  demonstrate that if the cavity length is selected properly, we can
  cross the point where the non-lasing state is becoming unstable,
  nevertheless obtaining good fundamental CML regime. That is, no need
  of a seed pulse is necessary anymore, in contrast to previous
  considerations \cite{kozlov1997self,kozlov1999self,
    kozlov2011obtaining,kozlov2013single,menyuk2009self,talukder2009analytical}. The
  resulting pulsed attractor is stabilized globally due to finite-size
  effects in the cavity. 
  We focus our attention on the gaseous media with $T_2$ in the range
  of nanoseconds, so that we have no need to approach single-cycle
  limit to overcome $T_2$, which significantly simplifies the required
  model.  Nevertheless, this scheme is very attractive as a source of
  picosecond pulses with high power and high repetition rate $>1$
  GHz. 

\section{The model}

The ring-cavity configuration considered in this article is shown in
\reffig{fig:2}a. Between the mirrors, only one of which is assumed
partially reflecting with the reflection coefficient $R$, and the
others are ``ideal'' for simplicity, the gain and the absorber
sections are placed. Both sections consist of resonant nonlinear
medium, tuned to the same frequency. The coupling to the field, namely
the dipole moment of active ``atoms'' is different for both
sections. The media and field are described in the two-level and
slowly-varying envelope approximations respectively \cite{Allen:book,klykanov01,klykanov02}:
\begin{gather}
\partial_t P(z,t)=-\dfrac{P(z,t)}{T_{2}(z)}+
\dfrac{d_{12}(z)}{2\hbar}\Delta\rho(z,t)A(z,t),
\label{eqPspm} \\
\partial_t \Delta\rho(z,t)=-\dfrac{\Delta\rho(z,t)-
\Delta\rho_{0}(z)}{T_{1}(z)}-\dfrac{d_{12}(z)}{2\hbar}F(z,t),
\label{eqRho} \\
\partial_t A(z,t)-
c\partial_z A(z,t)=
4\pi\omega_{12} d_{12}(z)N_{0}(z) P(z,t).
\label{eqApm} %\\
\end{gather}
Here $P(z,t)$ is the slowly-varying envelope of the non-diagonal
element of the density matrix describing the two-level atom in the
absorber (for $0<z<L_a$) and gain (for $L_a<z<L\equiv L_a+L_g$) sections in rotating
wave approximation; $\Delta \rho(z,t)$ is the difference of diagonal
elements of the density matrix (population difference per single atom)
in the gain and absorber sections, $A(z,t)$ is the lowly varying field
amplitude in the gain and absorber sections. $\omega_{12}=2\pi
c/\lambda_{12}$ is the transition frequency,
 $c$ is the speed of light  in vacuum, and $F(z,t)=A(z,t)P(z,t)$; The other parameters,
and their values in numerical simulations are given in the
Table~\ref{tb:1} and are characteristic for gases \cite{chatak2009optics}.
 The cavity length is $L=30$ cm
  (corresponding to round-trip time $\tau=1$ns), and the mirror
reflectively is $R=0.8$. The equations
  (\ref{eqPspm})-(\ref{eqApm}) are highly used in different physical
  situations describing resonant behavior of nonlinear media
  \cite{mccall1969self,kryukov70,Allen:book,babushkin98,babushkin00a, schuttler08,
    schulz-ruhtenberg10,arkhipov2013effects}.

%\begin{widetext}
%\begin{center}
\begin{table*}[t!]
\caption{\label{tb:1} Parameter values used in the numerical simulations}
\label{tab:par}%{}
%\begin{centering}
\begin{tabular}{| p{6cm}|l|l|}
\hline
Parameter & Gain $(L_a<z<L)$ & Absorber $(0<z<L_a)$  \tabularnewline
\hline
central transition wavelength  & $\lambda_{12} = 0.6$ $\mu m$ & $\lambda_{12} = 0.6$ $\mu m$    \tabularnewline
length of the medium & $L_{g} = 15$ cm & $L_a=15$ cm    \tabularnewline
concentration of two-level atoms, $N_{0}(z)=\ldots$ & $N_{0g}=4.0\cdot 10^{13}$ cm${^{-3}}$ & $N_{0a} = 0.8\cdot 10^{13}$ cm${^{-3}}$   \tabularnewline
transition dipole moment, $d_{12}(z)=\ldots$ & $d_{12g}$ = 0.5 Debye &  $d_{12a}$ = 1 Debye   \tabularnewline
population difference at equilibrium, $\Delta \rho_{0}(z)=\ldots$ & $\Delta \rho_{0g}= -1$ & $\Delta \rho_{0a} = 1$   \tabularnewline
population difference relaxation time, $T_{1}(z)=\ldots$ & $T_{1g}=6$ ns & $T_{1a}=6$ ns    \tabularnewline
polarization relaxation time, $T_{2}(z)=\ldots$ & $T_{2g}=0.5 $ ns & $T_{2a}=0.5$ ns   \tabularnewline
\hline
\end{tabular}
%\par
%\end{centering}
\end{table*}
%\end{center}
%\end{widetext}
%
We remark that, because we assume the gaseous media with $T_2$
  in the range of nanoseconds, we do not need to approach single-cycle
  pulse limit to achieve the coherent (SIT) regime. In our case the pulse duration will be $>1$ ps, thus,
  the dispersion of the cavity plays only the minor role. Besides,
  because the ratio of Rabi frequency to $\omega_{12}$ is $\ll
  10^{-3}$ in our simulations, the two-level approximation can be
  assumed as very good one.  

Important quantity characterizing the pulse propagation dynamics in
two level atoms is the pulse area $\Phi$:
\begin{equation}
\Phi(z,t) =
  \frac{d_{1,2}}{\hbar}\int_{-\infty}^{\infty}A(z,\tau)d\tau \approx  
\frac{d_{1,2}}{\hbar}\int_{t-4\tau_p}^{t+4\tau_p}A(z,\tau)d\tau,\label{eq:phi}
\end{equation}
where $\tau_p$ is the pulse duration; that is, we integrate in the
region around the pulse to get practically useful measure in the case
when other pulses are present.  

\section{Results of numerical simulations}

According to the original idea of the
coherent soliton mode locking
\cite{kozlov1997self,kozlov1999self,kozlov2011obtaining,menyuk2009self,
  talukder2009analytical,kozlov2013single}, the
dipole moments of the gain is two times smaller than in the absorber,
which ensures, together with the ``mixed'' character
  of the medium, that fluctuations in the near of the non-lasing state decay.
Such situation can be described from the point of view
  of nonlinear dynamics \cite{haken:book} in the following way: the stable pulsed regime exists
and is well separated in the phase space (see inset to Fig.~1c) from the non-lasing steady state, which is also
stable. That
  is, the system is attracted  to the regime,
in which vicinity it is located initially. Other regimes may exist but
they are irrelevant for our
consideration.
To achieve a stable pulsed regime in this situation one needs a
large-intensity ``perturbation'', that is, a seed pulse, which makes
the scheme more complicated from the practical point of view.

%%%
\begin{figure}[htbp!]
\begin{centering}
\includegraphics[width=0.5\textwidth]{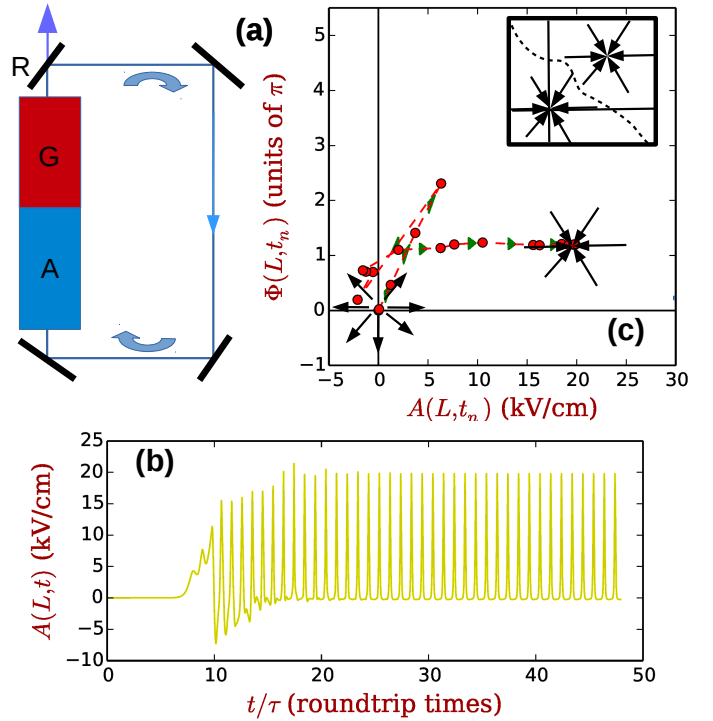}%%
\caption{ \label{fig:2} (a) The scheme of the ring-cavity laser with
  an absorber (A) and gain (G) sections. (b) The output field
  $A(L,t)$ evolution in dependence on time during a buildup of a
  stable pulsed regime from a vicinity of the non-lasing state. (c)
  The map in a plane of the pulse area $\Phi(L,t_n)$
    and field $A(L,t_n)$ in the time moments  $t_n=t_0+n\tau$, where
    $n=1,2\ldots$ and $t_0$ is a constant chosen to align $t_n$
  to the pulse maxima at
    large $t_n$. Black arrows show the attracting
  or repelling character of every
  steady-state.  Green arrows show the
    direction of the system evolution.  In the inset the corresponding dynamical
    picture for the configuration from \cite{kozlov1997self} is
  shown. The laser parameters are given in Table~\ref{tb:1}.
 %%%
}
\end{centering} 
\end{figure}

 %\cite{kozlov1997self}.
Although in the infinite-length, ``mixed'' medium the soliton will be
shortened until it reaches single-cycle regime, in the cavity geometry
with the limited cavity length the pulse duration achieves its
stationary value $\tau_p$, which is determined by the cavity
parameters. The order of magnitude of $\tau_p$, e.g. in absorber, can
be obtained from the requirement $\Phi\approx2\pi$ leading to
$A\tau_p\approx 2\pi\hbar/d$. Using also the energy balance in the
gain section, that is, assuming that the energy of the pulse produced
by the atoms $w\approx\hbar\omega_{12}L_gN_g$ is fully transferred to
the field, that is, $w\approx A^2\tau_p/8\pi c$, we will finally
obtain:
 %\cite{kozlov1997self}.
%
\begin{equation}
  \label{eq:taup}
 \tau_p\approx \pi c \hbar/(d_a^2\omega_{12}N_gL_g),
\end{equation}
which gives for the parameters we use the order of
  tens picoseconds.  We note that ~\refeq{eq:taup}
  gives only very raw approximation of the pulse duration. Better
  approximation is possible using the equations given in
  \cite{kozlov1997self,talukder2009analytical}.

We assume the lengths of the gain and absorber sections $L_g$ and
$L_a$, larger than the pulse length $c\tau_p$ $L_a,L_g> c\tau_p$.
Importantly, we also assume that the cavity is small enough to prevent
the population difference to relax to their equilibrium states, that
is $\tau \ll T_{1g},T_{1a}$.  The last condition is
crucial and makes the pulse propagating in the cavity essentially
non-solitonic as we will see later.
The important parameter $G_0$
determining the gain-loss balance for low-intensity fluctuations and thus the
stability of non-lasing state is defined as:
\begin{equation}
%G_0 = \ln{\left[R\exp{(L_g\alpha_a-L_a\alpha_a)}\right]},
G_0 = \ln{\left[R\exp{\left\{\int_0^L\alpha(z)dz\right\}}\right]},
\label{eq:g}
\end{equation}
where  $\alpha(z)=-2\pi d^2_{12}(z) \omega_{12}
  N_{0}(z)\Delta \rho(z)T_{2}(z)/c\hbar$ is the gain/loss coefficient for low-intensity
fluctuations.  We chose $N_a$ and $N_g$ (see Table~1) in such a way that $G_0>0$,
facilitating  growth of low-frequency fluctuations
in the vicinity of nonlasing state and thus making it unstable.

The parameters for our numerical simulations are shown in
Table~\ref{tb:1}. One can see that $G_0\approx9.0>0$, that is, the
non-lasing state is strongly unstable. We start the simulations from its
vicinity, assuming the field close to zero, the population differences
at their equilibrium values and the polarization to be zero, and track
the evolution of the field in the cavity with time.  An exemplary
trajectory is shown in \reffig{fig:2}b, where we have started from the
a short probe pulse of the the duration $\tau_{p0}=100$ ps
and initial pulse area is $\Phi_0=10^{-7}\pi$.

After a short but
irregular transient process up to $t/\tau\sim 20$ a stable regime with the pulse duration
$\tau_p=80$ ps is born, which is significantly smaller that the
relaxation time $T_{2}$, thus corroborating the coherent character of
the mode-locking regime in this case.

We tested numerically the achievability and stability of the
pulsed regime, preforming many simulations starting initial
conditions randomly selected every time in large range of pulse
shapes, with or without initial noise. All the trajectories were attracted to
the single stable steady-state identical to the shown in
\reffig{fig:2}b,c up to a time shift and sign of the output field.

%%%
%\begin{widetext}
\begin{figure}[tphb!]
  \begin{centering}  
    \includegraphics[width=0.5\textwidth]{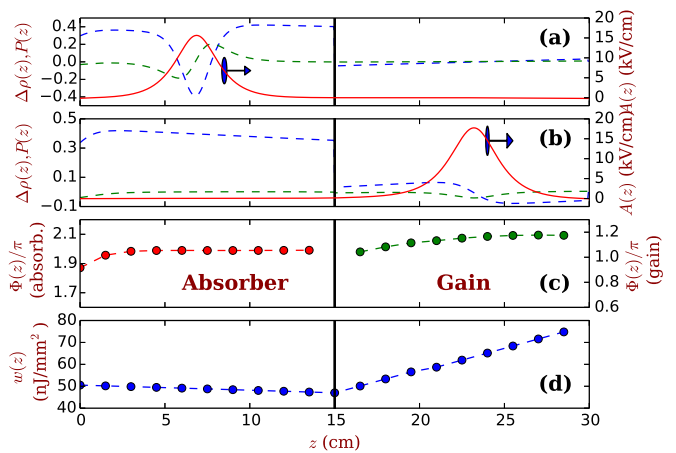}%%
\caption{ \label{fig:3} Pulse shapes in the amplifier (a) and absorber
  (b) sections in the stationary regime. The field amplitude $A(z)$
  (red line) as well as the polarization $P_s(z)$ (green dashed line)
  and population difference $\Delta \rho(z)$ (blue dashed line) are
  shown for the parameters in Table~\ref{tb:1}. Note the difference
  in sign in comparison to $N_{gain}$, $N_{absorber}$ in Fig.~1.  In (c)
  and (d), the pulse
  area $\Phi$  (c) and the pulse energy flux $w$ (d) in the gain and absorber sections are shown.
 %%%
}
\end{centering}
\end{figure}
%\end{widetext}

The details of the resulting steady-state regime are shown in
\reffig{fig:3}, where the pulse shapes in the absorber
\reffig{fig:3}(a) and amplifier \reffig{fig:3}(b) sections are shown,
together with the pulse area $\Phi$ in the cavity \reffig{fig:3}(c).
As on can see, the pulse area in absorber, initially (at $z=0$) being
reduced to around $1.9\pi$ by the losses at the mirror, approaches
quickly it's stable value $2\pi$. The population difference is
 positive before the pulse,
becomes  negative at its leading
edge, and at the trailing edge is restored to the initial value
demonstrating dynamics which is typical for $2\pi$-pulses.
In addition, despite of the lossy character of the absorbing medium, the
  pulse energy does not decreases upon the propagation in the absorber, as it is shown in
  \reffig{fig:3}(d), which is also typical for
  $2\pi$ pulses.
In the gain section, the pulse area is close to $\pi$, as seen from
\reffig{fig:3}(c). Typically for such pulses, the population inversion
changes its sign after the pulse propagation. Also, as
  seen from \reffig{fig:3}(d), the pulse ``eats'' the energy from the
  gain medium, so its full energy increases.  Then, the energy
  obtained in the gain section is radiated thought the mirror at
  $z=L$. 

Importantly,  the population
difference before the pulse is far from its equilibrium value, and
rather close to zero (although remains negative), that is, the medium
is in slightly amplifying regime before the pulse. After the pulse,
the gain medium is switched to slightly absorbing regime. As the
relaxation time $T_{1g}\gg \tau$, the gain section has no time to
achieve its stationary value after the pulse.  In contrast, the
population difference in the absorber remains comparable to its
equilibrium value. As a result, the gain-loss balance for the small
perturbations becomes negative, protecting the pulse against
destroying. Namely, $G_0$ in the such pulsed regime
  (obtained by integrating everywhere outside of the pulse in
  Eq.~\ref{eq:g}) becomes negative, in particular for parameters in
  Fig.~2 $G_0\approx-1.0$.

%The pulse duration $\tau_p$ for the regime in FIg.~2 is around $80$ ps

\begin{figure}[htbp!]
  \begin{centering}
    \includegraphics[width=0.5\textwidth]{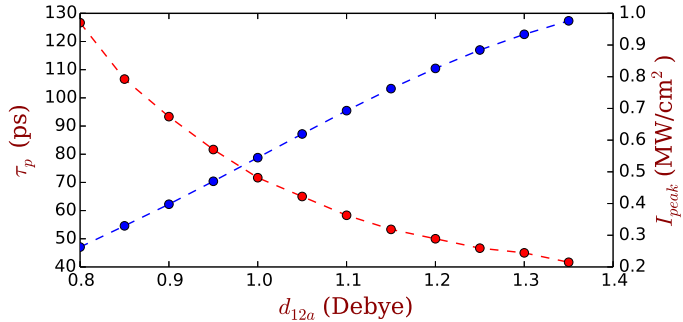}%%
\caption{ \label{fig:4} The pulse duration $\tau_p$ (red
    line) and the peak
  intensity $I_{\mathrm{peak}}$ (blue line) in dependence on the $d_{12a}$ in the
  stationary regime, changing the concentration in absorber section in
  such a way that $G_0$  remains the same, and keeping all other
  parameters as in Table~\ref{tb:1}
 %%%
}
\end{centering}
\end{figure}

This dynamics is rather typical for wide range of the parameters as
far as the conditions $L_a,L_g>c\tau_p$ and $L\ll cT_{1g},cT_{1a}$,
$G_0>0$ (and is not very large, $G_0\lessapprox20$)
 are satisfied. For instance, in
\reffig{fig:4} the dependence of pulse duration and the peak intensity
on the dipole moment of absorber $d_{12a}$ is shown, assuming the
other parameters except $N_{0a}$ being unchanged,
the later is scaled such that $G_0$
   remains constant. The pulse durations  are far from the
  single-cycle limit but nevertheless  are significantly smaller than
  the phase coherence time.
% We conclude that the ``dynamical protection'' phenomena
% described above is rather robust and depends only on the above
% mentioned relation between $\tau_p$, $L_a,L_g$ and $T_1$.

Now let us briefly consider what happens if we cross these cavity
length limits in various directions. If $L_g$ approaches $cT_{1g}$, the
population difference is able to recover before the pulse returns. In
the other words, now more than one such pulse can propagate inside the
cavity. In this situation, according to our simulations, the number of
pulses in the cavity is indeed larger than one.
The pulses typically interact with each other, leading to a
complicated dynamics, so that the perfect mode-locking does not
  take place anymore. An example of such regime is presented in Fig.~4 (blue
  line) for $L=50$ cm.   In the case when $L$ decreases and becomes

comparable with $\tau_p$, the soliton-like structure can not anymore
exist, and either (a spatially inhomogeneous)
steady-state, or periodic regime with the low
  modulations depth appears, as shown in Fig.~4 (green
  and red lines) for $L=18$ cm and $L=20$ cm, correspondingly.

\begin{figure}[htbp!]
  \begin{centering}
    \includegraphics[width=0.5\textwidth]{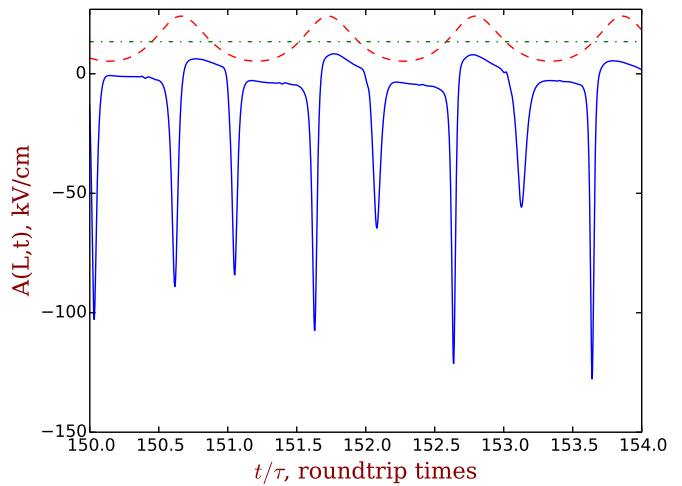}%%
\caption{ \label{fig:4n} The field dynamics at the laser output $A(L,t)$ in
  dependence on time for different cavity lengths $L=50$ cm (blue solid
  line),  $L=20$ cm (red dashed line), $L=18$ cm (green
  dash-dotted line). The other parameters are as in Fig.~1. 
}
\end{centering}
\end{figure}

\section{Discussion and conclusion}

In conclusion, we have shown numerically that a stable self-starting
CML allowing pulse durations significantly smaller than $T_2$ can be
obtained in a laser with absorber and gain sections separated in
space.  No seed pulse and no complicated, difficult to realize
artificial geometry considered in previous works on CML is
needed. Rather trivial ring-cavity two-section laser with properly
selected parameters makes the stable mode-locked regime being
automatically self-started from the non-lasing state. This is a
significant advantages over the previous approaches to CML which were
all non-self-starting.  To achieve self-starting the non-lasing state
must be unstable and, in addition, the cavity round-trip time $\tau$
must be chosen to be significantly smaller than the population
difference relaxation time $T_{1g}$ but still significantly larger
than the pulse duration.
In this case, during the onset of generation process, initial
fluctuations start to grow until the laser enters significantly nonlinear
and even coherent regime with $\Phi\sim 1$.  This initiates a
chaotic behaviour with, typicaly, a train of several irregular pulses
propagating in the cavity. This is a point where our stabilization
mechanism starts to work. After a passage of the strongest pulse the
population inversion starts to restore on the time scale of $T_{1g}$.
If the cavity lengths is selected properly, this pulse makes the full
roundtrip and returns back just at the point when the restoration is
(almost) completed, whereas the other pulses ``fill'' smaller gain and thus are
depleted. In this way, the mode-locking regime with only one pulse per
roundtrip time is stabilized.

% We note that the mechanism of CML self-starting considered here starts
% to work already in the strongly nonlinear coherent regime. 
In the other words, the mechanism considered here stabilizes the
fundamental modelocking regime against the multiple pulses per
roundtrip time.
%%%
In this respect, we are also free from the necessity to consider
development of the fluctuation dynamics at the lasing threshold, as it is
typically made for self-starting modelocking schemes in the cases when
the fundamental modelocking is developing from the
non-lasing steady state
\cite{ippen90,krausz91,haus91,herrmann93,krausz93,chen95,kalosha99a,gordon02,komarov02,soto-crespo02,gordon06}. In
the same way it is clear that the mechanism described here gives the
modelocking with the probability one.
%, in contrast to modelocking
%regimes starting from the background fluctuations which develop with
%some finite probability. 
We remark also that, because of the instability of the off state,
the regime considered here also can not be described as a temporal
cavity soliton \cite{marconi14}.
  
%We remark that we 
We considered here a case of a diluted gas with
  typical $T_2$ in the range of nanoseconds, which makes the
  description of the problem rather simple even if the pulse duration
  $\tau\ll T_2$. In our case $\tau$ is in the range of tens of
  picoseconds. In particular, dispersion of the cavity elements play
  no role and can be neglected. Also, because the Rabi frequency is
  also very small in comparison to $\omega_{12}$, the two-level
  approximation considered here is also rather good. %one.
 
  Nevertheless, even in the present case, far from the single-cycle limit and
  with rather low density of the active atoms, CML-based scheme
  demonstrates impressive pulse repetition rate in the range of GHz
  and high power density. We remark that our preliminary simulations
  show the possibility of such stable CML regime for large range of
  parameters (cavity length, dipole moments etc), for
  instance, typical for semiconductor lasers. Nevertheless, to
  describe such system more comprehensive modeling is needed which
  will be subject of our future studies.

   R.~Arkhipov would like to acknowledge the support of EU FP7 ITN
   PROPHET, Grant No. 264687.  Authors also thank Dr.~I.~A.~Chekhonin
   for helpful discussions.

\section*{References}

%\bibliography{optics0,optics}

\end{document}